# Scattering properties of *PT*- symmetric layered periodic structures

O.V. Shramkova, G.P. Tsironis

Crete Center for Quantum Complexity and Nanotechnology, Department of Physics, University of Crete
P.O. Box 2208, 71003  Heraklion, Greece

**Abstract**
The optical properties of *PT*-symmetric periodic stack of the layers with balanced loss and gain are examined. We demonstrate that tunnelling phenomenon in periodic structures is connected with excitation of surface waves at the boundaries separating gain and loss regions within each unit cell and tunnelling conditions for periodic stack can be reduced to the conditions for one period. Alternatively, it is shown that coherent perfect absorber laser states are mediated by excitation of surface modes localised at all internal boundaries of the structure. The effects of the structure parameters and angle and direction of incidence on the resonant phenomena and spontaneous symmetry breaking transition are determined. It is shown that structure periodicity significantly increases the number of resonant phenomena, especially in stacks with high real and imaginary parts of dielectric permittivity of the layers. Gaussian beam dynamics in this type of structure is examined. The beam splitting in *PT*-symmetric periodic structure is observed.

1. **Introduction**

Interest to artificial electromagnetic media such as metamaterials and photonic crystals has steadily grown during the last few years. Search and the development of the new types of the artificial media with novel functional capabilities across the whole spectrum of frequencies ranging from radiowaves to visible will have significant impact on advances in the development of future communications, computing systems, and many other applications of artificial media. The discovery of parity-time (*PT*)-symmetric media has shown significant potential for advancing toward the goal of new metamaterial design.

The recent theoretical investigations of Bender and Boettcher [1] have indicated that even non-Hermitian Hamiltonians can exhibit entirely real spectra as long as they respect the conditions of *PT*-symmetry. The effects associated with the *PT*- symmetric systems have been comprehensively investigated in quantum mechanics [2] and optics [3]-[4] during the last years. In the context of optics, *PT*-symmetric systems have attracted much attention both theoretically and experimentally [4]-[7] . It was demonstrated that these materials can exhibit several exotic features, including unidirectional invisibility [8], coherent perfect absorption [9], nonreciprocity of light propagation [10]-[12], beam refraction [10], multistability [13] and various extraordinary nonlinear effects [14]-[17].

The *PT* -related concepts can be realized in artificial optical materials that rely on balanced gain and loss regions. In this framework, *PT*-symmetry demands that the complex refractive index obeys the condition $n(\vec{r}) = n^*(-\vec{r})$. However, the latter is only a necessary condition for *PT* symmetry, because the transition to a complex spectrum, which is called *PT*- symmetry breaking, appears upon the increase of the strength of imaginary part of $n(\vec{r})$. Spontaneous *PT*- symmetry breaking has been experimentally observed in optics [10], [18] and has been studied theoretically for different models [19]-[21].

To date, most of the studies in optical realizations of *PT*-symmetric media have connected with investigation of stacked layers and films [22]-[26]. The planar structures containing dielectric layers with balanced gain and loss  are of particular interest because on one hand they are compatible with the existing fabrication technologies and, on the other hand, their basic models provide deep insight in the main features and mechanisms of the scattering. The existence of transmission resonances in which the reflectance vanishes only for waves incident from one side of the structure, which we refer to as anisotropic transmission resonances (ATRs), unidirectional invisibility phenomenon (as a special case of these ATRs) and  *PT*- breaking transitions in the spectrum of 1D *PT* -symmetric photonic heterostructure were recently discussed in [22]. The optical medium, consisting of a uniform index grating with two homogeneous and symmetric gain and loss regions, that realizes a *PT*- symmetric coherent perfect absorber

(CPA) laser was introduced in [23]. It was demonstrated that such systems can behave simultaneously as a laser oscillator, emitting outgoing coherent waves, and as a CPA, absorbing incoming coherent waves. The effect of non-Hermiticity parameters (loss-gain balance) on the dispersion and scattering properties of a metamaterial, composed of a periodically stacked five-layer plasmonic waveguide, was investigated in [26]. It is necessary to note that investigation of periodical *PT*- symmetric structures was primary limited by consideration of Bragg gratings under normally incident wave illumination [8][27],[28]. The spectral features of nonreciprocal waveguide Bragg gratings with careful adjustment of the parameters corresponding to the index and gain/loss gratings were considered in [27]. It was demonstrated that the scattering from Bragg structures with *PT*-symmetric refractive index distribution can become unidirectionally invisible near the spontaneous *PT*-symmetry breaking point [8].

The aim of the present work is to explore the influence of the periodicity and individual layer parameters onto the optical properties of *PT*-symmetric stack of binary dielectric layers under oblique incident plane-wave illumination. This paper is organized as follows. The problem statement and main analytical derivations are outlined in Section 2 and elaborated in Appendix. The features of *PT*- symmetric structures composed of dielectric multilayers are illustrated by the results of numerical simulations and discussed in Section 3. We show that tunnelling and CPA-laser phenomena mediated by the excitation of surface wave at the boundaries between the layers with balanced loss and gain can occur in periodic *PT*-symmetric multilayers. The spontaneous symmetry breaking phenomena in periodic stack and influence of periodicity on the exceptional points of scattering matrix eigenvalues are discussed. The resonant phenomena and effect of structure parameters on the reflectivity and transmittivity of *PT*-symmetric stack are analysed. The main findings are summarised in Conclusion.

## 2. Problem statement and main equations

Let us consider a finite periodic structure formed by two alternating slabs of identical thickness $d$, with complex-conjugate dielectric permittivities $\varepsilon = \varepsilon' - i\varepsilon''$ and $\varepsilon^* = \varepsilon' + i\varepsilon''$ ($\varepsilon'$ and $\varepsilon''$ are positive) corresponding to balanced gain and loss regions, respectively. The system is *PT*-symmetric about $z=0$. The geometry of the problem is shown in Fig.1. The stack of the layers of the total thickness $L=2Nd$ ($N$ is the number of structure periods) is embedded in a homogeneous medium with dielectric permittivity $\varepsilon_a$ at $z \leq -L/2$ and $z \geq L/2$. Since the layers are assumed isotropic in the *x-y* plane, the TE and TM polarised waves with the fields independent of the *y*-coordinate ($\partial/\partial y = 0$) can be analysed separately. Only the TM waves with the field components $E_x, E_z, H_y$ are discussed in the rest of the paper.

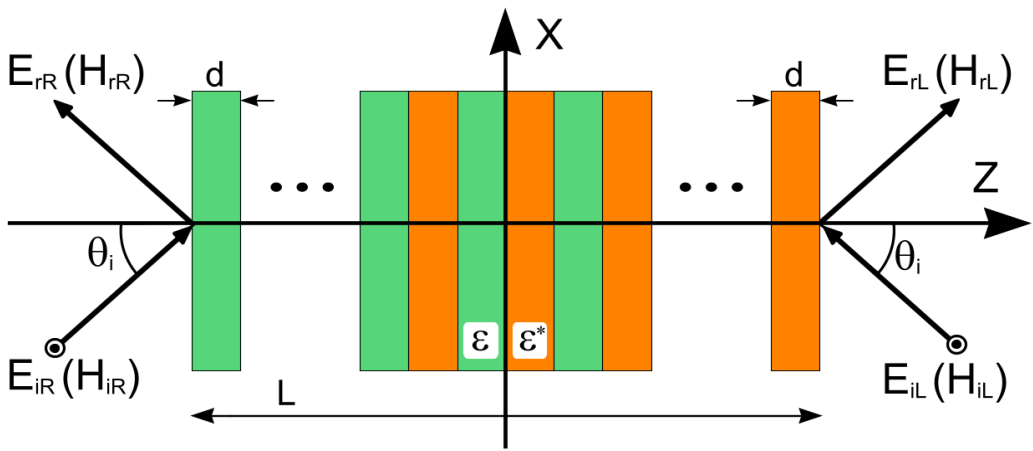

Fig.1. Geometry of the problem.

The magnetic field of TM wave in the surrounding homogeneous medium can be expressed as

$$H_y(\omega,x,z) = e^{-i\omega t + i k_x x} \begin{cases} A e^{i k_{za}(z+L/2)} + B e^{-i k_{za}(z+L/2)}, & z \leq -L/2 \\ C e^{i k_{za}(z-L/2)} + D e^{-i k_{za}(z-L/2)}, & z \geq L/2 \end{cases}, \qquad (1)$$

where $k_x = k_0 \sqrt{\varepsilon_a} \sin\theta_i$ is the transverse wave number; $k_0 = \omega/c$, $c$ is the speed of light; $\theta_i$ is the angle of wave incidence; $k_{za} = k_0 \sqrt{\varepsilon_a} \cos\theta_i$ is the longitudinal wave number outside the stack; $A$, $B$, $C$ and $D$ are amplitudes of the forward and backward propagating waves outside the stack. Satisfying the continuity conditions for the tangential field components at the stack outer interfaces we obtain

$$\begin{pmatrix} A \\ B \end{pmatrix} = \frac{1}{2} \begin{pmatrix} 1 & \frac{\omega}{c} \frac{\varepsilon_a}{k_{za}(\omega)} \\ 1 & -\frac{\omega}{c} \frac{\varepsilon_a}{k_{za}(\omega)} \end{pmatrix} \widehat{M}(\omega) \begin{pmatrix} 1 & 1 \\ -\frac{c}{\omega} \frac{k_{za}(\omega)}{\varepsilon_a} & \frac{c}{\omega} \frac{k_{za}(\omega)}{\varepsilon_a} \end{pmatrix} \begin{pmatrix} D \\ C \end{pmatrix}, \qquad (2)$$

where $\widehat{M}(\omega)$ is the transfer matrix of the whole stack obtained by transfer matrix method (TMM). The transfer matrix $\widehat{M}(\omega)$ of the finite periodic structure containing $N$ periods can be expressed in terms of the transfer matrix $\widehat{m}(\omega) = \widehat{m}_{L1}(\omega) \widehat{m}_{L2}(\omega)$ of a single period using Abeles theorem [29]: $\widehat{M}(\omega) = (\widehat{m}(\omega))^N$, where $\widehat{m}_{Lj}(\omega)$ is the transfer matrix of the $j^{th}$ layer.

Making use of (2) we can deduce the $S$-matrix in terms of the stack transfer matrix elements. The $S$-matrix is defined by

$$\begin{pmatrix} B \\ C \end{pmatrix} = \widehat{S} \begin{pmatrix} A \\ D \end{pmatrix} = \begin{pmatrix} R^{(L)}(\omega) & T(\omega) \\ T(\omega) & R^{(R)}(\omega) \end{pmatrix} \begin{pmatrix} A \\ D \end{pmatrix}, \qquad (3)$$

where $R^{(L,R)}(\omega)$ are stack reflection coefficients for wave incident from the left and right, $T(\omega)$ is the transmission coefficient (it is the same for left and right incidence). The close form expressions for reflection/transmission coefficients are provided in Appendix. It is necessary to note that from the unitarity of $S$-matrix we get the conservation relation [22] $\left| |T(\omega)|^2 - 1 \right| = \left| R^{(L)}(\omega) R^{(R)}(\omega) \right|$. The scattering matrix in definition (3) measures the breaking of $PT$ symmetry. The eigenvalues $\lambda_{1,2}$ of $S$-matrix must have reciprocal modules or condition $|\lambda_1 \lambda_2| = 1$ should be satisfied. In the $PT$-symmetric phase these eigenvalues are unimodular and for symmetric phases $|\lambda_1| = |\lambda_2| = 1$. As it was demonstrated in the literature [22], a 1D $PT$-symmetric structures can undergo spontaneous symmetry-breaking transitions in the eigenvalues and eigenvectors of its $S$-matrix when an abrupt phase transition to a complex eigenspectrum takes place. In the $PT$-symmetry breaking points both eigenvalues meet and bifurcate, and the broken phases correspond to $|\lambda_1| = 1/|\lambda_2| > 1$.

An alternative definition of the scattering matrix can be used for the determination of unidirectional invisibility points and takes the form

$$\begin{pmatrix} D \\ A \end{pmatrix} = \widehat{S}_c \begin{pmatrix} B \\ C \end{pmatrix} = \begin{pmatrix} T(\omega) & R^{(L)}(\omega) \\ R^{(R)}(\omega) & T(\omega) \end{pmatrix} \begin{pmatrix} B \\ C \end{pmatrix}. \qquad (4)$$

The ATR [8], [22] for which $T=1$ and one of the reflection coefficients vanishes, can be observed in $PT$-symmetric periodic structures. This phenomenon can be referred as unidirectional reflectivity and is associated with the exceptional points for the eigenvalues $\lambda_{1c,2c}$ of the $S_c$-matrix corresponding to the tunnelling conditions for left and right incidence.

Let us consider the dispersion properties of infinite dielectric *PT*-symmetric structure. Using the TMM method and applying the Floquet theorem, which takes into account the periodicity of the structure, we obtain the following well-known dispersion relation [30]

$$\cos 2\bar{k}d = \frac{m_{11}+m_{22}}{2} = \cos k_{z1}d \cos k_{z2}d - \frac{1}{2}\left(\frac{\varepsilon_1 k_{z2}}{\varepsilon_2 k_{z1}} + \frac{\varepsilon_2 k_{z1}}{\varepsilon_1 k_{z2}}\right)\sin k_{z1}d \sin k_{z2}d, \qquad (5)$$

where $\bar{k}$ is the Bloch wavenumber (averaged wavenumber, describing the periodicity of the structure); $k_{zj} = \sqrt{k_0^2 \varepsilon_j - k_x^2}$ ($j$=1,2) is the longitudinal wave numbers in the layers; $\varepsilon_1 = \varepsilon$ and $\varepsilon_2 = \varepsilon^*$ are dielectric permittivities of the layers; $m_{11}$ and $m_{22}$ are elements of the unit cell transfer matrix. We investigate the dispersion relation for a finestratified medium, i.e., we assume that $k_{zj}d \ll 1$. In this case, the Bloch wavenumber $\bar{k} = k_z^{(fs)}$ is the transverse wavenumber of a uniaxial medium and the dispersion relation for a fine-stratified dielectric structure has the form

$$\frac{k_x^2}{\varepsilon_{zz}} + \frac{\left(k_z^{(fs)}\right)^2}{\varepsilon_{xx}} = \frac{\omega^2}{c^2}, \qquad (6)$$

where effective components of the tensor of dielectric permittivity for *PT*-symmetric structure can be written as

$$\varepsilon_{xx} = \frac{\varepsilon_1 d + \varepsilon_2 d}{2d} = \varepsilon'$$
$$\varepsilon_{zz} = 2d\left(\frac{d}{\varepsilon_1} + \frac{d}{\varepsilon_2}\right)^{-1} = \frac{\varepsilon'^2 + \varepsilon''^2}{\varepsilon'} \qquad (7)$$

Since Eq. (6) relates the three quantities $k_z^{(fs)}, k_x$ and $\omega$, it is convenient to regard it as describing either the contours of constant frequency. According to dispersion relation, a contour of constant frequency is a second-order curve. The effective components of dielectric permittivity tensor (7) are positive ($\varepsilon_{xx}, \varepsilon_{zz} > 0$) and the contour is an ellipse for all values of frequency, as in the case of a uniaxial dielectric crystal. It is necessary to note that for simulated below ENZ materials with $\varepsilon'' > \varepsilon'$ ($\varepsilon' = 0.0001$, $\varepsilon'' = 0.001$) the ratio between ellipse axes will be equal to 10 and *PT*-symmetric structure can be treated as anisotropic dielectric. At equal $\varepsilon'$ and $\varepsilon''$ the ratio between ellipse axes will be reduced to $\sqrt{2}$. For the low level of loss and gain ($\varepsilon'' \ll \varepsilon'$), the contour is a circle and fine-stratified *PT*-symmetric structure can be considered as isotropic structure.

## 3. Properties and Mechanisms of Resonant Scattering Phenomena

To elucidate the influence of the periodicity and constitutive parameters of the layers on the features of *PT*-symmetric structures, the characteristics of the TM wave scattering by the stacks of periodically sequenced binary dielectric layers have been analysed numerically. Illustrative examples of the simulation results for the periodic multilayers are discussed below in comparison with the respective *PT*-symmetric bilayers.

### 3.1 Tunneling Conditions

Recently, the metamaterials with ENZ permittivities have attracted great attention owing to the specified wave interaction properties. It was shown that working in the ENZ regime with vanishingly small real part of dielectric permittivity we can obtain a markedly visible tunnelling phenomenon at low levels of gain [24] in *PT*-symmetric bilayer. The tunnelling phenomena in ENZ bilayers with balanced loss and gain are mediated by the excitation of a surface wave at the interface separating the gain and loss regions. Authors of the mentioned work have determined the tunnelling conditions and analysed three regimes of operation connected with gain/loss level.

Let us consider the tunnelling phenomena that can occur in the ENZ *PT*-symmetric periodic structures. The analysis of the expressions for reflection coefficients has enabled an insight in the fundamental properties of the scattering by periodic stack of the layers illuminated by obliquely incident plane wave. In our work the analytical study of expressions for reflection coefficient $R^{(L,R)}(\omega)$ has revealed that total transmission conditions for the periodic layered structure can be reduced to the conditions for one period and conditions for Wolf-Bragg resonances connected with the total thickness of the stack and number of periods (see Appendix). Finally, we get that for *PT*-symmetric periodic stack the tunnelling/zero-reflection conditions can be fulfilled if for sufficiently thick layers the angle of wave incidence approaches the critical angle

$$\theta_c = \arcsin\sqrt{\frac{\varepsilon'^2 + \varepsilon''^2}{2\varepsilon'}}. \tag{8}$$

This expression for the critical angle coincides with the expression obtained for *PT*-symmetric ENZ bilayers in [24]. Moreover, similar to the case for two slabs the plane wave incident at $\theta_i > \theta_c$ can excite the surface waves at the gain-loss interfaces (Fig.2) with transversal propagation constant $k_x = k_0\sqrt{\varepsilon_a \frac{\varepsilon'^2 + \varepsilon''^2}{2\varepsilon'}}$. The peaks of distribution at the interfaces between the layers in the period confirm this fact. Magnetic field distributions presented in Fig.2 correspond to the transmission resonance at frequency $\omega = 2.947 \times 10^{13}$ $s^{-1}$ for wave incident from the left.

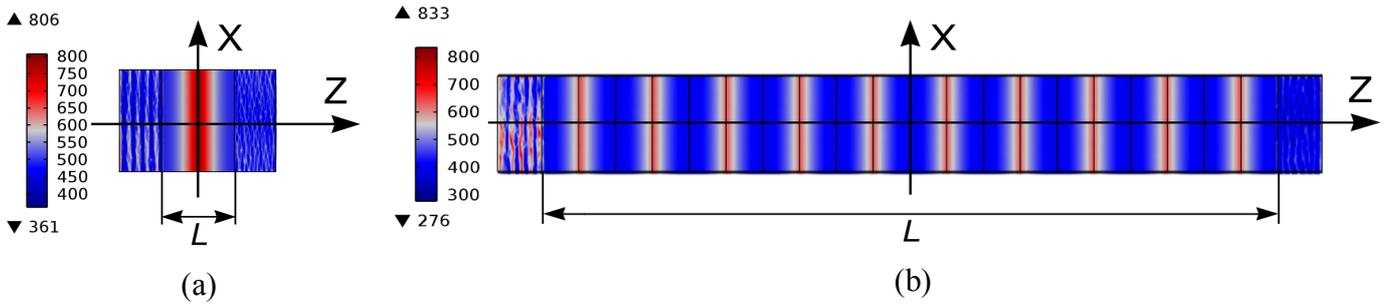

(a)          (b)

Fig.2. Distribution of the magnetic-field magnitude $|H_y|$ (the units of measurement are [A/m]) for incidence from left at $\theta_i = 5°$, $\omega = 2.947 \times 10^{13}$ $s^{-1}$, $\varepsilon' = 0.0001$, $\varepsilon'' = 0.001$, $d = 125$ μm; (a)- $N=1$, (b)- $N=10$ (The numerical simulation was carried out by software package COMSOL Multiphysics).

It is necessary to note that as in the case of *PT*-symmetric bilayers [24] the tunnelling phenomena in periodic *PT*-symmetric structures are determined by the relationship between bilayer permittivities, direction of incidence and electric thickness of the layers. All regimes of operation connected with gain/loss level considered for *PT*-symmetric ENZ bilayers are valid for binary periodic *PT*-symmetric stacks.

### 3.2 PT-symmetry breaking

The earlier studies of *PT*-symmetric systems reveal the existence of spontaneous symmetry breaking transition to a phase with a complex eigenspectrum [22]. The stack overall thickness may have profound influence on these phenomena. This can be the result of the increased number of unit cells in the stack or variations in the thicknesses of the constituent layers. As mentioned in the Section 2, the eigenvalues of the scattering *S*- and $S_c$-matrixes identify the exceptional points at which the symmetry-breaking transition and ATR occur. To illustrate the effect of the stack overall thickness, the frequency dependences of *S*-matrix eigenvalues for 1D *PT*-symmetric periodic stack of the ENZ layers with $\varepsilon' = 0.0001$, $\varepsilon'' = 0.001$ for angle of incidence $\theta_i$ close to the critical angle $\theta_c$ are displayed in Fig.3. The numerical examination of dependences for $\lambda_{1,2}$ shows that with increase of *N* (increase of total thickness of the stack) symmetry breaking occurs at lower frequency. In Fig.3b for *N*=50 the transition points reenter the symmetric phase at the frequencies close to Bragg resonances. The numerical analysis of the symmetry breaking transition for 2 different ENZ stacks with the same overall stack thicknesses $L$=250 μm but increasing number of the periods and thinner constituent layers is presented in Fig.3a and Fig.3c. It can be observed that, for increased number of unit cells in the stack and constant total thickness of the stack, the transition tends to occur at higher frequencies. The latter effect is confirmed by the analytical study for the fine-stratified periodic structure with layer thicknesses much less than the wavelength of electromagnetic wave. As it was demonstrated before, such structures can be considered as anisotropic dielectric with real effective components of dielectric permittivity tensor. So, the symmetry-breaking transitions will not be observed in fine-stratified $PT$-symmetric structures. At the same time, it is necessary to note that form and values of dependencies are determined by the number of periods (see Fig.3b and Fig.3c for $N$=50), and frequency shift depends on the thickness of the layers.

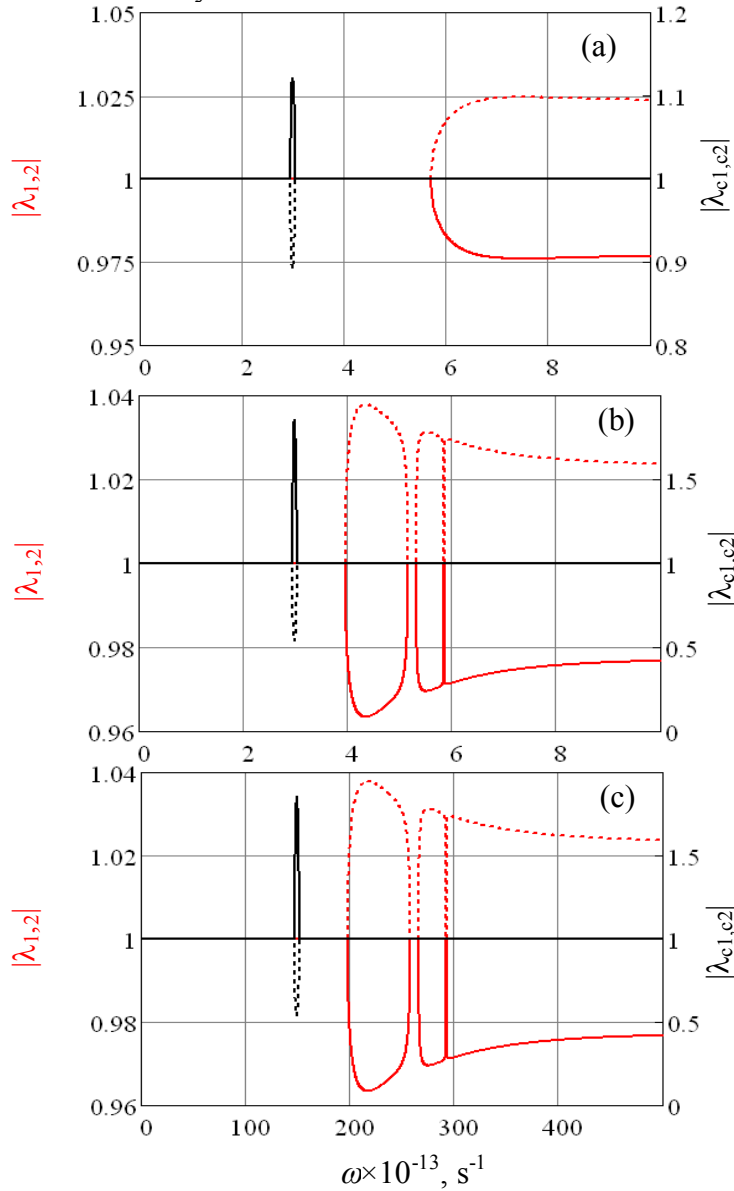

Fig.3. Modulus of eigenvalues of $S$- and $S_c$- matrixes for 1D $PT$-symmetric periodic stack of the layers as a function of frequency at $\varepsilon' = 0.0001$, $\varepsilon'' = 0.001$, $\theta_i = 5°$; (a) - $d$=125 μm, $N$=1; (b) - $d$=125 μm, $N$=50; (c) - $d$=2.5 μm, $N$=50. Red curves correspond to the eigenvalues $\lambda_{1,2}$, black curves are for $\lambda_{c1,c2}$; solid and dashed curves of the same color correspond to the different eigenvalues of the same scattering matrix. Note the different scales in the graphs.

Let us note that for increasing angles of incidence, the transition in periodic structures occurs at lower frequencies and the magnitudes of scattering matrix eigenvalues decrease. It was determined that at bigger $\theta_i$ for structure with $N$=50, $S$-matrix has a single transition to the broken symmetry phase connected with the absence of Bragg resonances. The exceptional points of $S_c$-matrix can be observed only for angles of incidence approaching the critical angle.

The reflectance/transmittance of electromagnetic waves and corresponding $S$-matrix eigenvalues are essentially dependent on the permittivities of the constituent binary layers. Therefore the effect of the dielectric permittivity has been assessed first to determine the contributions of $\varepsilon'$ and $\varepsilon''$ magnitudes to the breaking of $PT$- symmetry and existence of singular points in the broken-symmetry phase. In order to evaluate this phenomenon, the eigenvalues of two scattering matrixes have been simulated at higher real part of dielectric permittivity $\varepsilon'$ and higher level of loss and gain $\varepsilon''$. The modulus of the eigenvalues of the scattering matrixes simulated at the variable number $N$ of the unit cells are displayed in Fig. 4 for the stacks with $\varepsilon' = \varepsilon'' = 0.1$. Examination of frequency dependencies for $|\lambda_{1,2}|$ and $|\lambda_{1c,2c}|$ shows that with increase of real and imaginary parts of dielectric permittivities the ATR points move to the broken-symmetry region. The exceptional points for which one eigenvalue of scattering matrix tends to infinity and second goes to zero correspond to the CPA laser points. For a one-period structure (Fig.4a,c) at $\omega = 1.49 \times 10^{13}$ $s^{-1}$, the system is very near the CPA laser threshold point where we can observe pole and zero of the scattering $S$- and $S_c$- matrixes. The abrupt changing of the $S_c$- matrix eigenvalues in the singular point (Fig.4c) is connected with the changing of the $R^{(R)}$ reflection coefficient phase from $\pi$ to $-\pi$. However, the number $N$ of stacked unit cells has strong impact on the $S$-matrix eigenvalues. Just like in the case of ENZ structure, the $S$-matrix transition moves to lower frequencies with an increase of $N$. But in some cases (see for example Fig.4b for $N=10$) the frequency phase transition will be suppressed by Bragg resonance. It is noteworthy that the additional singular points arise in the thicker stacks (Fig.4a,c). The Bragg resonances in stack create additional modulation of the $|\lambda_{1,2}|$ and $|\lambda_{1c,2c}|$ magnitudes and frequency shift of singular points.

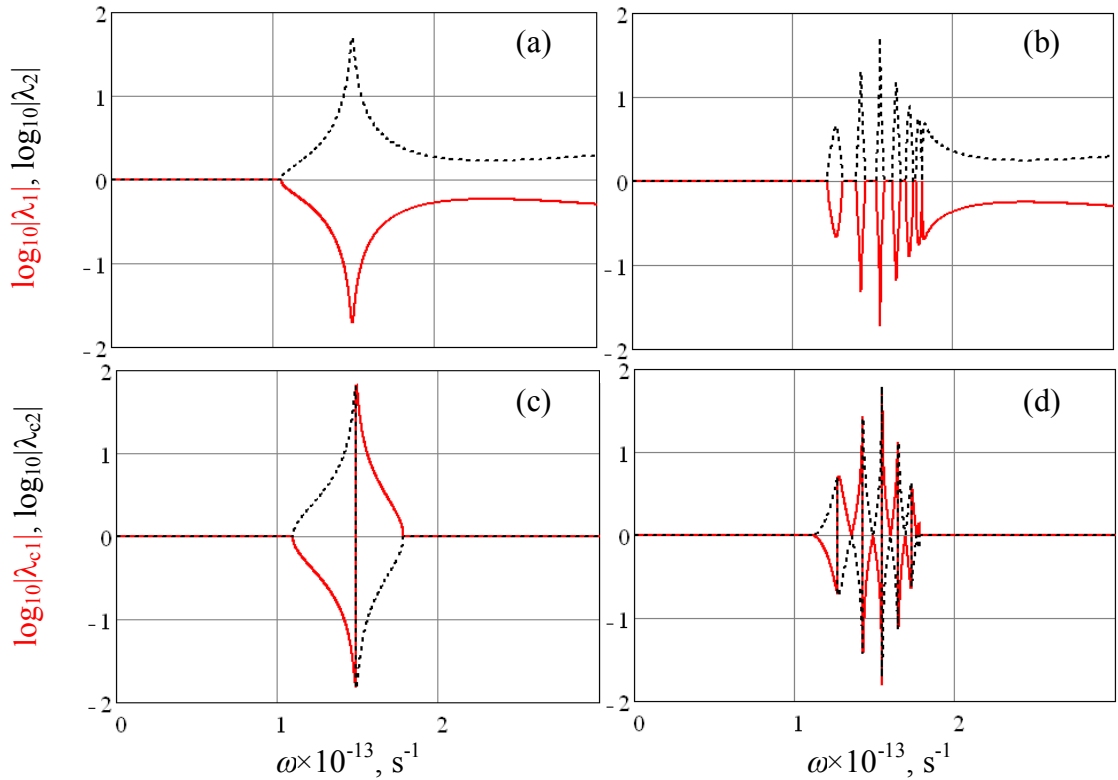

Fig.4. Logarithm of the modulus of eigenvalues of $S$- (a,b) and $S_c$- matrixes (c,d) for 1D $PT$-symmetric periodic stack of the layers as a function of frequency at $\varepsilon' = 0.1$, $\varepsilon'' = 0.1$, $d=0.125$ µm, $\theta_i = 5°$, (a,c) – $N=1$, (b,d) – $N=10$; solid red and dashed black curves correspond to the different eigenvalues of the scattering matrix.

The earlier studies of lasing modes in 1D cavities with spatially inhomogeneous gain profile have demonstrated that CPA laser states are connected with surface modes localised at the gain-loss boundary [31]. The physical origin of the surface waves was identified as the transmission resonances of the gain-free region. To gain deeper insight the nature of CPA-lasing phenomena in *PT*-symmetric stacks of the layers, it is necessary to consider the field distribution along z direction at the frequency corresponding to singular point of scattering matrix. The magnetic field distribution for the binary dielectric layers with the default parameters as in Fig. 4 and at CPA laser points is displayed in Fig. 5. It can be seen that the field is localised at all interfaces between the layers. As the result, the amplification of reflected and transmitted modes can be observed.

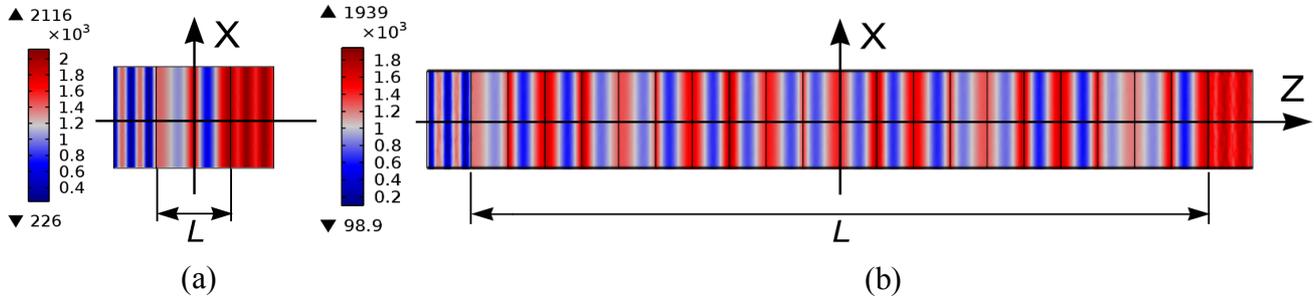

(a) (b)

Fig.5. Distribution of the magnetic-field magnitude $|H_y|$ (the units of measurement are [A/m]) for incidence from left at $\theta_i = 5°$, $\varepsilon' = 0.1$, $\varepsilon'' = 0.1$, $d=125$ μm; (a)- $\omega = 1.49\times10^{13}$ $s^{-1}$, $N$=1; (b)- $\omega = 1.55\times10^{13}$ $s^{-1}$, $N$=10 (The numerical simulation was carried out by software package COMSOL Multiphysics).

To assess the effect of the real part of dielectric permittivity $\varepsilon'$ on spontaneous symmetry breaking and resonant phenomena, magnitudes $|\lambda_{1,2}|$ and $|\lambda_{1c,2c}|$ have been simulated at increased values of $\varepsilon'$. It was demonstrated that number of singular and exceptional points of scattering matrixes was not determined only by the number of the unit cells. This number rises with dielectric permittivity of the layers. Let us note that at high values of $\varepsilon'$ the angle of incidence has minor effect on the symmetry breaking and singular points.

*3.3 TM wave reflection/refraction by PT-symmetric stacks*

To illustrate the effect of the stack periodicity, the reflectivity for both left and right incidence and the transmittivity of TM waves incident at slant angle $\theta_i = 5°$ along with phases of corresponding coefficients are displayed in Fig.6 for ENZ bilayers (Figs.6a,b for one unit cell) and periodic stack of alternating ENZ layers (Figs.6c,d for 50 unit cells). The constituent layer parameters for these 2 stacks are the same as in Figs.3a,b. It is necessary to note that Figs.6 b,d are zoomed in on the ATRs. Two very closely spaced ATRs (one for left incidence and one for right) correspond to frequencies $\omega = 2.947\times10^{13}$ $s^{-1}$ and $\omega = 3.054\times10^{13}$ $s^{-1}$. As it was mentioned before, these frequencies do not depend on the number of structure periods. The magnitudes of the reflectivity/transmittivity will be changed close to the points of ATR. In Figs.6a,c these resonances are marked by vertical dotted lines. All other reflectionless transmission resonances (Bragg resonances) in Fig.6c are attributed to the increase of the stack overall thickness. At these resonance points structure reflectionless for both directions of incidence. The number of points corresponding to Bragg resonances depends on the number of periods $N$. The amplification of reflected waves is connected with a constructive interaction between the forward- and backward-propagating waves. Figs. 6b,d show that the phases $\phi^{(L)}$ and $\phi^{(R)}$ of corresponding reflection coefficients $R^{(L)}$ and $R^{(R)}$ jump at each ATR and Bragg resonances. At the same time at Bragg resonances the phase of transmission coefficient is equal to zero ($\phi^{(T)}$=0 at $\omega = 3.41\times10^{13}$ $s^{-1}$, see Fig.6d) as in a case of ATRs or we get jump of transmission phase.

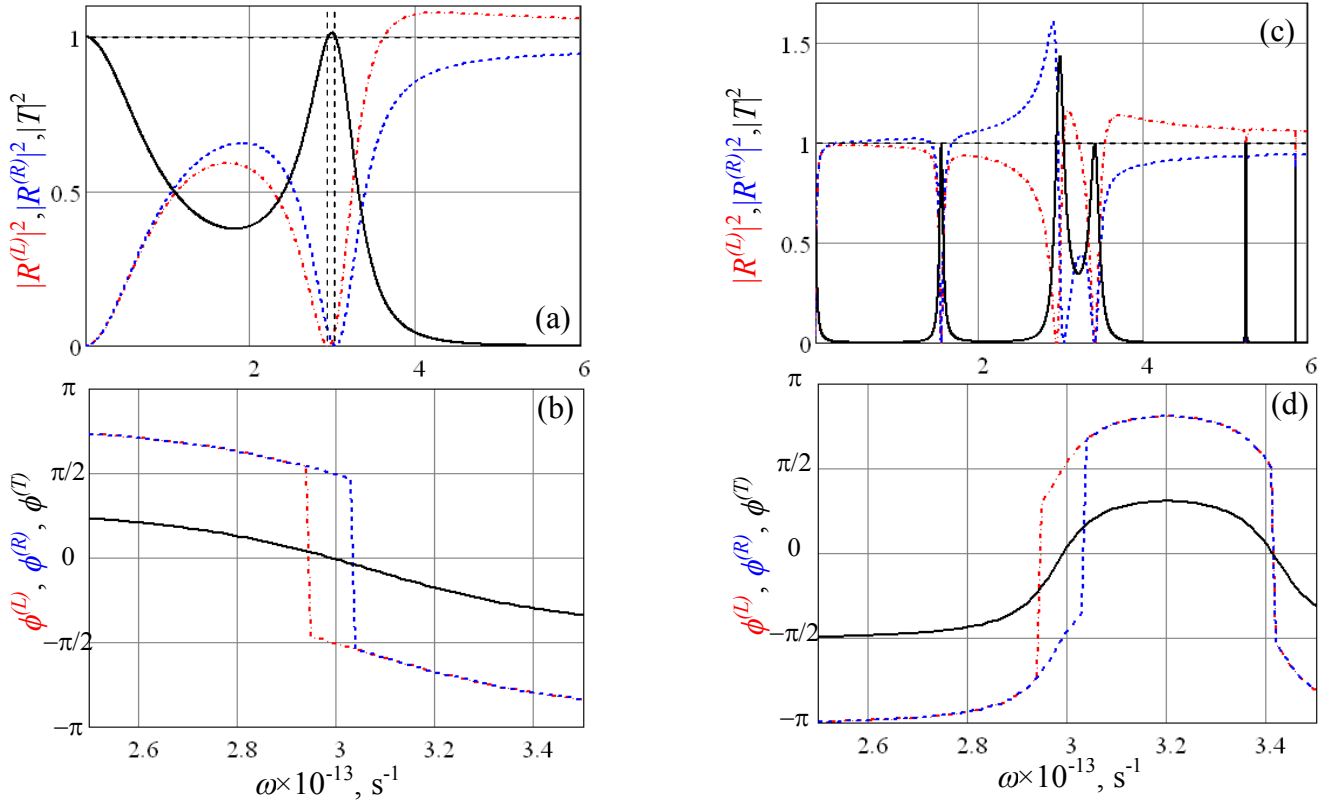

Fig.6. (a,c) - TM wave reflectance ($|R^{(L)}|^2$ – dash-dot red curve, $|R^{(R)}|^2$ – dashed blue curve) and transmittance (solid black curve), (b,d) - phases of $R^{(L)}$ ($\phi^{(L)}$- dash-dot red curve), $R^{(R)}$ ($\phi^{(R)}$- dashed blue curve) and T ($\phi^{(T)}$- solid black curve) for *PT*-symmetric stack with $\varepsilon' = 0.0001$, $\varepsilon'' = 0.001$, $d$=125 μm, (a,b,c) - *N*=1, (d,e,f) - *N*=50, illuminated at the incidence angle $\theta_i = 5°$. Note the different scales in the graphs.

To estimate the effect of dielectric permittivity magnitude, the structures with the same parameters as in Fig. 4 have been simulated. The reflectance of TM waves incident at slant angle on the stack of the layers with $\varepsilon' = 0.1$, $\varepsilon'' = 0.1$ is displayed in Fig.7. As in a previous case, the considerable differences in reflectivity/transmittivity of *PT*-symmetric bilayer (Figs.7a,b) and periodic stack with *N*=10 (Figs.7c,d) are evident, especially at low frequencies, and connected with Bragg resonances for thicker structure. It is evident that at opposite slant angles of incidence reflectivity exhibits noticeable nonreciprocity and we can observe 2 ATRs for right incidense and wave amplification for the left one. Peaks of reflectivity and transmittivity at $\omega = 1.49 \times 10^{13}\ s^{-1}$ for bilayer structure and multiple peaks corresponding to singular points in Fig.4b for *N*=10 are the CPA laser resonant points. In these points we can observe the jump of reflection phases $\phi^{(L)}$, the reflection phases $\phi^{(R)}$ are equal to 0 and the transmission phases $\phi^{(T)}$ are equal to $\pm\frac{\pi}{2}$. The charater of phases corresponding to other resonant phenomena was discussed above and will not be changed.

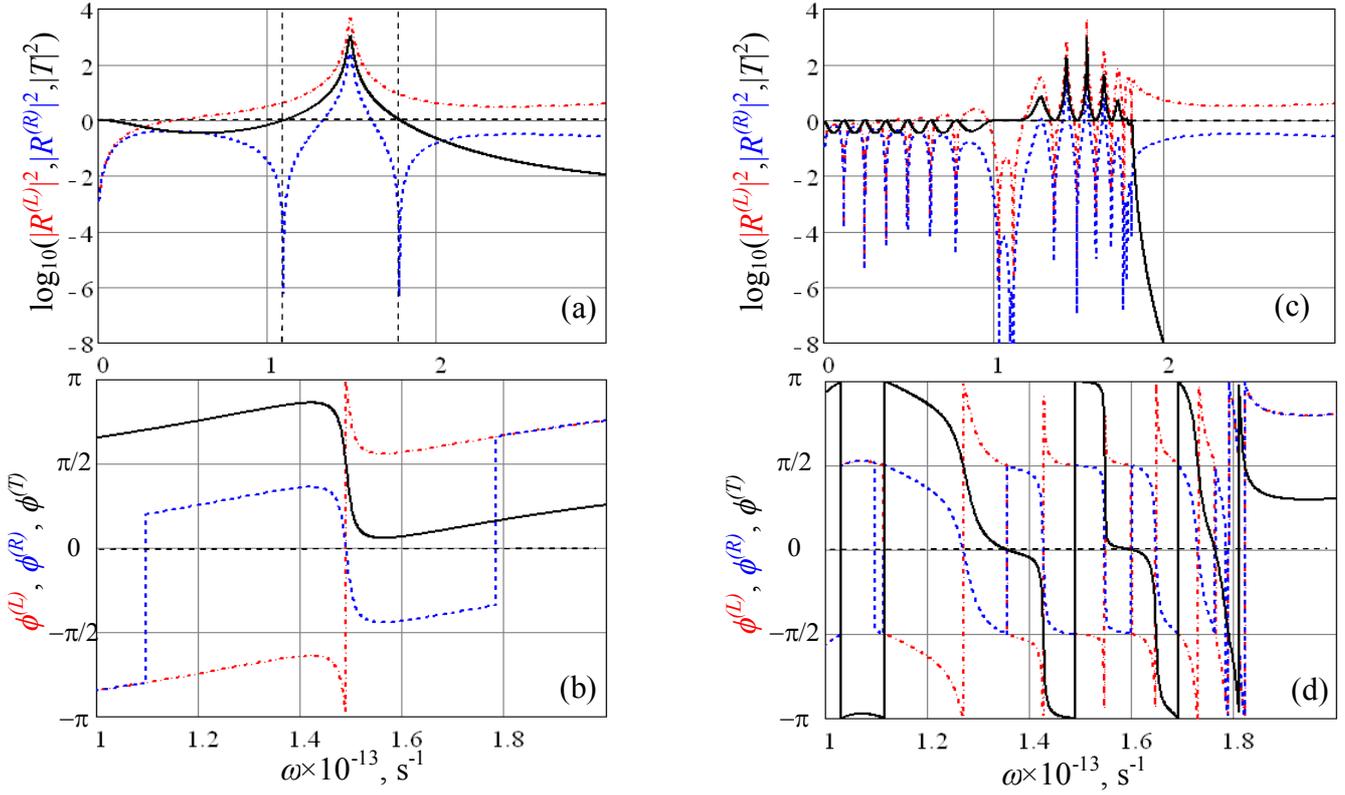

Fig.7. (a,c) - TM wave reflectance ($|R^{(L)}|^2$ – dash-dot red curve, $|R^{(R)}|^2$ – dashed blue curve) and transmittance (solid black curve), (b,d) - phases of $R^{(L)}$ ($\phi^{(L)}$- dash-dot red curve), $R^{(R)}$ ($\phi^{(R)}$- dashed blue curve) and T ($\phi^{(T)}$- solid black curve) for PT-symmetric stack with $\varepsilon' = 0.1$, $\varepsilon'' = 0.1$, $d=125$ μm, (a,b) - N=1, (c,d) - N=10, illuminated at the incidence angle $\theta_i = 5°$. Note the different scales in the graphs.

As it was mentioned before, the number of ATRs and singular points extremely depends on dielectric permittivity. Therefore, when real part of dielectric permittivity is increased to $\varepsilon' = 1.1$, the number of ATR points for right incidence rises, as shown in Fig.8. In the case of left incidence at $\omega > 7.6\times 10^{13}$ $s^{-1}$ (Fig.8a for N=1) system works as amplifier. With increase of stack number (Fig.8b for N =10) system can be considered as amplifier in wider frequency range due to the symmetry breaking transition at lower frequency. Moreover, we get huge number of CPA laser points.

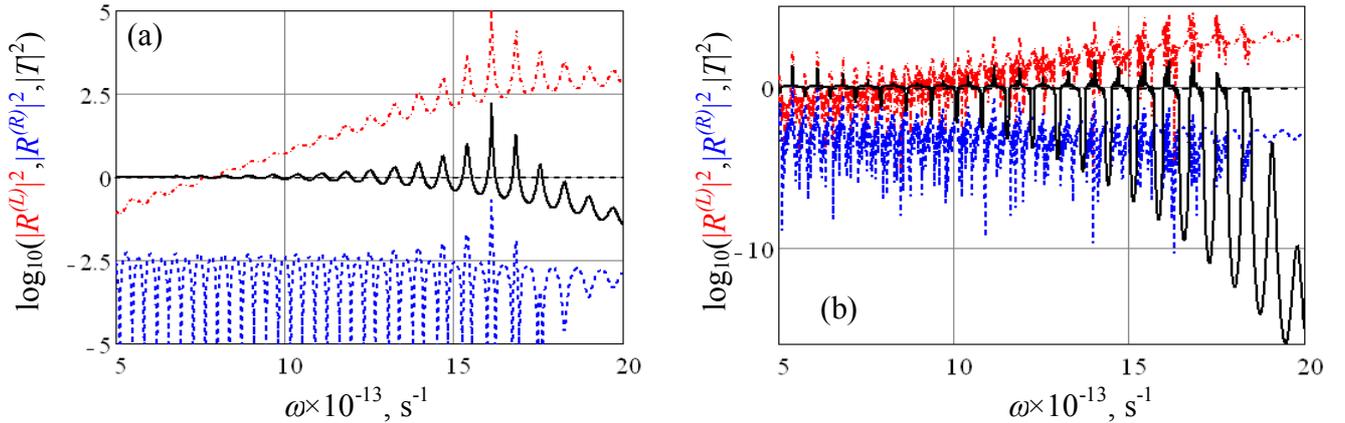

Fig.8. TM wave reflectance ($|R^{(L)}|^2$ – dash-dot red curve, $|R^{(R)}|^2$ – dashed blue curve) and transmittance (solid black curve), for PT-symmetric stack with $\varepsilon' = 1.1$, $\varepsilon'' = 0.1$, $d=125$ μm, (a) - N=1, (b) - N=10, illuminated at the incidence angle $\theta_i = 5°$.

*3.4 Gaussian beam dynamics in PT-symmetric periodic structures*

In the following we will focus on the Gaussian laser beam propagation in *PT*-symmetric stacks. To investigate the Gaussian beam evolution in *PT*-symmetric periodic structures the electric and magnetic field distributions for 1D *PT*-symmetric periodic stack of the ENZ layers with $\varepsilon' = 0.0001$, $\varepsilon'' = 0.001$ under normal incidence at frequencies of transition to the broken symmetry phase are displayed in Fig. 9. The numerical simulation was carried out by software package COMSOL Multiphysics. Comparison of Figs. 9(a,b) and 9(c,d) for left and right normal incidence demonstrates strong nonreciprocal response of the structure. The amplification of reflected beam and low intensity of transmitted beam at left incidence are observed. At the same time, the intensities of reflected and refracted beams at right incidence are almost equal. It is necessary to note that for normal incidence the symmetrical beam splitting inside the structure takes place. The amplitudes of intensity oscillations are not constant and maximal amplitude of oscillations corresponds to x=0. Moreover, inside the structure the magnetic field intensity is close to zero. Such properties of laser beam propagating in *PT*-symmetric stacks are directly related to the power flow oscillations [10] and interference of reflected/refracted waves in the layers of the structure. Similar analysis of the field distribution has been performed for the periodic stacks containing *N*=50 unit cells at frequency close to *PT*-symmetry threshold (Fig.9e). We can observe the periodic undamped oscillations of intensity along the direction of beam propagation.

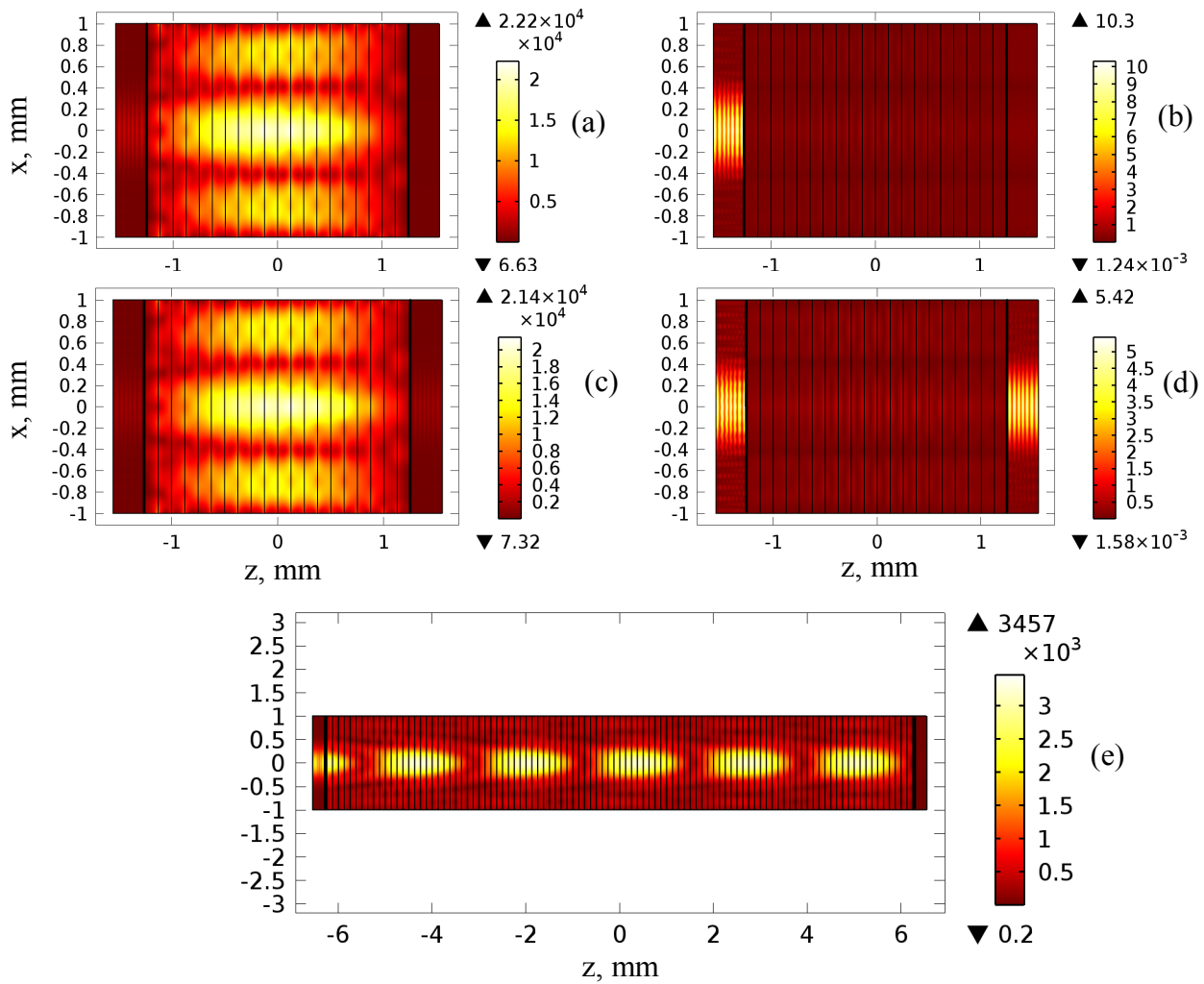

Fig.9. Distribution of the electric- (a,c,e - the units of measurement are [V/m]) and magnetic-field (b,d - the units of measurement are [A/m]) norm at $\varepsilon' = 0.0001$, $\varepsilon'' = 0.001$, $d$=125 μm for normal incidence from left (a,b) – *N*=10, $\omega = 7.31 \times 10^{13}$ $s^{-1}$, (e) - *N*=50, $\omega = 4.97 \times 10^{13}$ $s^{-1}$, and right (c,d) - *N*=10, $\omega = 7.31 \times 10^{13}$ $s^{-1}$.

Figs.10a,b show electric- and magnetic-field norm distribution for *PT*-symmetric periodic stack of binary dielectric layers with the default parameters as in Fig. 4(b) at frequency close to CPA lasing point for normal left incidence. As in a case of ENZ structure, the beam splitting is observed. The amplification

and widening of reflected and transmitted beams are demonstrated. The effect of the beam incidence angle on the field distribution is displayed in Figs.10c,d at $\theta_i = 5°$, $N=10$. It can be observed that for slant beam incidence the symmetry of field distribution about x=0 will be broken.

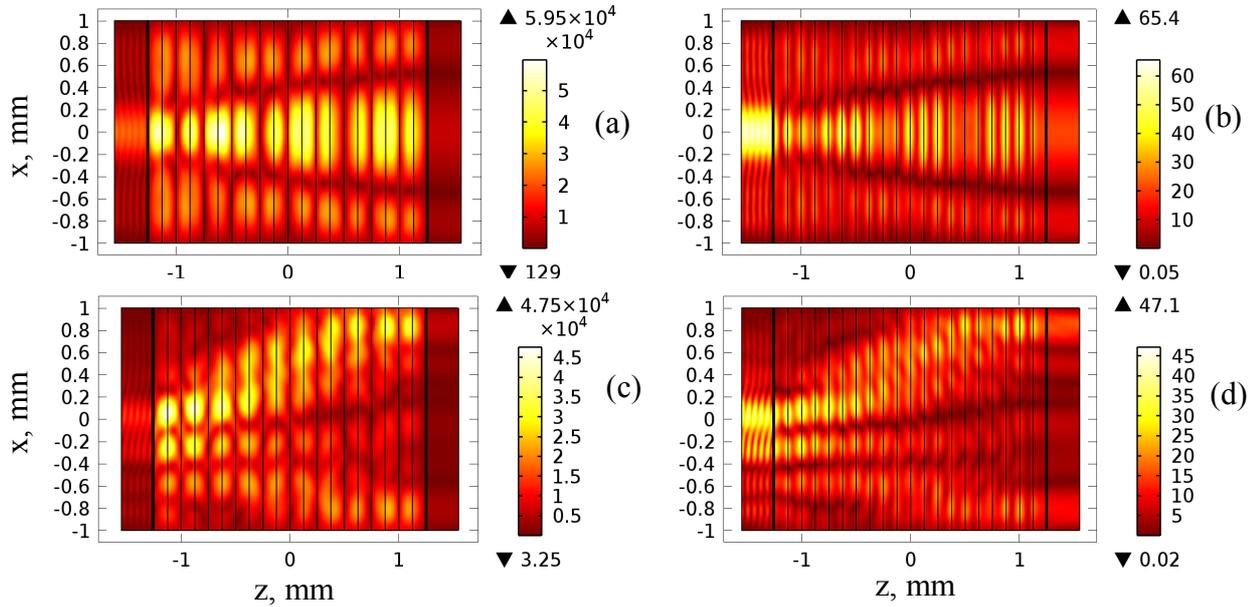

Fig.10. Distribution of the electric- (a,c - the units of measurement are [V/m]) and magnetic-field (b,d - the units of measurement are [A/m]) norm at $\varepsilon' = \varepsilon'' = 0.1$, $d=125$ μm, $N=10$, $\omega = 1.6 \times 10^{13}$ $s^{-1}$ for normal incidence from left (a,b) and right (c,d).

## 4. Conclusion

The spectral features of *PT*-symmetric periodic stack of binary dielectric layers characterised by balances loss and gain have been explored. Using the scattering matrix formalism, we have analysed the scattering properties and symmetry breaking transitions in *PT*- symmetric periodic structure. We have determined the parameter configuration for resonant phenomena. The main mechanisms and resonant scattering properties have been illustrated by the simulation results. It has been shown that ATRs and CPA laser states are connected with surface mode excitation at internal boundaries of the stack. It has been demonstrated that the resonant phenomena and spontaneous symmetry breaking are strongly influenced by the constitutive and geometrical parameters of the layers and wave angles of incidence. The performed parametric study has shown that the number of singular and exceptional points of scattering matrixes corresponding to resonant scattering phenomena is primary determined by the number of the unit cells and dielectric permittivity of the layers. The non-reciprocal behaviour of Gaussian beam is illustrated by the simulation results. The beam splitting in *PT*-symmetric structures is observed.

**Acknowledgment**
The research work was partially supported by the European Union Seventh Framework Program (FP7-REGPOT-2012-2013-1) under grant agreement No. 316165

**Appendix**

The reflection and transmission coefficients are obtained by imposing the continuity conditions for the tangential field components at the layer interfaces and can be represented in the form:

$$R^{(L)}(\omega) = \frac{M_{11}(\omega) + \frac{c}{\omega}\frac{k_{za}(\omega)}{\varepsilon_a}M_{12}(\omega) - \frac{\omega}{c}\frac{\varepsilon_a}{k_{za}(\omega)}M_{21}(\omega) - M_{22}(\omega)}{M_{11}(\omega) + \frac{c}{\omega}\frac{k_{za}(\omega)}{\varepsilon_a}M_{12}(\omega) + \frac{\omega}{c}\frac{\varepsilon_a}{k_{za}(\omega)}M_{21}(\omega) + M_{22}(\omega)}$$

$$R^{(R)}(\omega) = \frac{-M_{11}(\omega) - \frac{\omega}{c}\frac{\varepsilon_a}{k_{za}(\omega)}M_{21}(\omega) + \frac{c}{\omega}\frac{k_{za}(\omega)}{\varepsilon_a}M_{12}(\omega) + M_{22}(\omega)}{M_{11}(\omega) + \frac{\omega}{c}\frac{\varepsilon_a}{k_{za}(\omega)}M_{21}(\omega) + \frac{c}{\omega}\frac{k_{za}(\omega)}{\varepsilon_a}M_{12}(\omega) + M_{22}(\omega)} \quad (A1)$$

$$T(\omega) = \frac{2}{M_{11}(\omega) + \frac{c}{\omega}\frac{k_{za}(\omega)}{\varepsilon_a}M_{12}(\omega) + \frac{\omega}{c}\frac{\varepsilon_a}{k_{za}(\omega)}M_{21}(\omega) + M_{22}(\omega)}.$$

Using the Abeles theory [29] and terms of the transfer matrix $\hat{m}(\omega)$ [30] of a single period the components of transfer matrix of the whole stack can be written as

$$M_{11,22} = m_{11,22}\frac{\sin 2N\bar{k}d}{\sin 2\bar{k}d} - \frac{\sin 2(N-1)\bar{k}d}{\sin 2\bar{k}d},$$
$$M_{12,21} = m_{12,21}\frac{\sin 2N\bar{k}d}{\sin 2\bar{k}d}. \quad (A2)$$

where Bloch wavenumber $\bar{k}$ is determined in the text.

Let us determine the conditions of the full passage of electromagnetic waves, for which the reflectance is equal to zero. We can show that reflection coefficients are equal to zero if

$$F_1 \pm F_2 = 0, \quad (A3)$$

where "+" corresponds to $R^{(L)} = 0$, "-" refers to $R^{(R)} = 0$, and

$$F_1 = -M_{11} + M_{22}$$
$$F_2 = -\frac{\omega}{c}\frac{\varepsilon_a}{k_{za}(\omega)}M_{21} + \frac{c}{\omega}\frac{k_{za}(\omega)}{\varepsilon_a}M_{12}. \quad (A4)$$

As the result, we get that first of all full transmittance occurs at Wolf-Bragg resonances of Bloch waves in the finite periodic structures. At these resonances $|R^{(L,R)}(\omega)| = 0$ as the stack overall thickness equals an integer number of Bloch half-waves, that is, $2N\bar{k}d = \pi q$, $q = 0, \pm 1, \pm 2, \dots$. Moreover after unsophisticated analytical manipulations, we obtain that

- $F_1 = 0$ at $m_{11} - m_{22} = 0$;
- $F_2 = 0$ at $-\frac{\omega}{c}\frac{\varepsilon_a}{k_{za}(\omega)}m_{21} + \frac{c}{\omega}\frac{k_{za}(\omega)}{\varepsilon_a}m_{12} = 0$.

The above observables demonstrate that zero-reflection conditions for periodic stack can be reduced to the conditions for one period.